\documentclass[twocolumn]{aastex63}

\received{XX, 2020}
\revised{YY, 2020}
\accepted{June 3, 2020}
\submitjournal{ApJ}

\shorttitle{The space density of $M_{1450}\sim -22.5$ AGNs at $z\sim 5.5$}
\shortauthors{Grazian et al.}
\begin{document}

\title{
On the AGN nature of two UV bright sources at $z_{spec}\sim 5.5$ in the
CANDELS fields: an update of the AGN space density at $M_{1450}\sim -22.5$
}

\correspondingauthor{Andrea Grazian}
\email{andrea.grazian@inaf.it}

\author[0000-0002-5688-0663]{A. Grazian}
\affiliation{INAF--Osservatorio Astronomico di Padova,
Vicolo dell'Osservatorio 5, I-35122, Padova, Italy}

\author[0000-0003-0734-1273]{E. Giallongo}
\affiliation{INAF--Osservatorio Astronomico di Roma, via Frascati 33,
I-00078, Monteporzio, Italy}

\author[0000-0002-4031-4157]{F. Fiore}
\affiliation{INAF--Osservatorio Astronomico di Trieste, via G.B. Tiepolo 11,
I-34131, Trieste, Italy}

\author[0000-0003-4432-5037]{K. Boutsia}
\affiliation{Carnegie Observatories, Las Campanas Observatory,
Casilla 601, La Serena, Chile}

\author[0000-0002-2115-1137]{F. Civano}
\affiliation{Harvard-Smithsonian Center for Astrophysics, 60 Garden
Street, Cambridge, MA 02138, USA}
%Yale Center for Astronomy and Astrophysics, 260 Whitney
%Avenue, New Haven, CT 06520, USA

\author[0000-0002-2115-5234]{S. Cristiani}
\affiliation{INAF--Osservatorio Astronomico di Trieste, via G.B. Tiepolo 11,
I-34131, Trieste, Italy}

\author[0000-0002-6830-9093]{G. Cupani}
\affiliation{INAF--Osservatorio Astronomico di Trieste, via G.B. Tiepolo 11,
I-34131, Trieste, Italy}

\author[0000-0001-5414-5131]{M. Dickinson}
\affiliation{National Optical Astronomy Observatory, Tucson, AZ 85719, USA}

\author[0000-0003-4744-0188]{F. Fontanot}
\affiliation{INAF--Osservatorio Astronomico di Trieste, via G.B. Tiepolo 11,
I-34131, Trieste, Italy}

\author[0000-0002-4096-2680]{N. Menci}
\affiliation{INAF--Osservatorio Astronomico di Roma, via Frascati 33,
I-00078, Monteporzio, Italy}

\author[0000-0002-9948-3916]{M. Romano}
\affiliation{Dipartimento di Fisica e Astronomia, Universit\`a di Padova,
Vicolo dell'Osservatorio 3, I-35122, Padova, Italy}
\affiliation{INAF--Osservatorio Astronomico di Padova,
Vicolo dell'Osservatorio 5, I-35122, Padova, Italy}

%% Note that the \and command from previous versions of AASTeX is now
%% depreciated in this version as it is no longer necessary. AASTeX 
%% automatically takes care of all commas and "and"s between authors names.

%% AASTeX 6.3 has the new \collaboration and \nocollaboration commands to
%% provide the collaboration status of a group of authors. These commands 
%% can be used either before or after the list of corresponding authors. The
%% argument for \collaboration is the collaboration identifier. Authors are
%% encouraged to surround collaboration identifiers with ()s. The 
%% \nocollaboration command takes no argument and exists to indicate that
%% the nearby authors are not part of surrounding collaborations.

%% Mark off the abstract in the ``abstract'' environment. 
\begin{abstract}
It is a widespread opinion that hydrogen reionization is mainly driven
by primeval star-forming galaxies, with a minor role of high-z active
galactic nuclei. Recent observations, however, challenge this notion,
indicating a number of issues related to a galaxy-driven reionization
scenario. We provide here an updated assessment of the space density
of relatively faint ($M_{1450}\sim -22.5$) AGNs at $z_{spec}\sim 5.5$
in order to improve the estimate of the photo-ionization rate
contribution from accreting super massive black holes. Exploiting
deep UV rest-frame ground-based spectra collected at the Very Large
Telescope on the CANDELS/GOODS-South field and deep Chandra X-ray
images in the CANDELS/GOODS-North and EGS areas, we find two
relatively bright ($M_{1450}\sim -22.5$) AGNs at $z_{spec}\sim
5.5$. We derive an AGN space density of $\Phi=1.29\times 10^{-6}
cMpc^{-3}$ at $z\sim 5.5$ and M$_{1450}\sim -22.5$ by simply dividing
their observed number by the cosmological volume in the range
$5.0<z<6.1$. Our estimate does not consider corrections for
incompleteness, therefore it represents a lower limit, although
uncertainties due to cosmic variance can still be significant. This
value supports a high space density of AGNs at $z>5$, in contrast with
previous claims mostly based on standard color selection, possibly
affected by significant incompleteness. Our estimate for the AGN
photo-ionization rate at $z\sim 5.5$ is in agreement with the observed
values at similar redshifts, which are needed to keep the
intergalactic medium highly ionized. Upcoming JWST and giant ground
based telescopes observations will
improve the study of high-z AGNs and their contribution to the
reionization of the Universe.
\end{abstract}

%% Keywords should appear after the \end{abstract} command. 
%% See the online documentation for the full list of available subject
%% keywords and the rules for their use.
\keywords{Active galactic nuclei (16) --- X-ray active galactic nuclei (2035)
--- Reionization (1383) --- Surveys (1671)}

%% From the front matter, we move on to the body of the paper.
%% Sections are demarcated by \section and \subsection, respectively.
%% Observe the use of the LaTeX \label
%% command after the \subsection to give a symbolic KEY to the
%% subsection for cross-referencing in a \ref command.
%% You can use LaTeX's \ref and \label commands to keep track of
%% cross-references to sections, equations, tables, and figures.
%% That way, if you change the order of any elements, LaTeX will
%% automatically renumber them.
%%
%% We recommend that authors also use the natbib \citep
%% and \citet commands to identify citations.  The citations are
%% tied to the reference list via symbolic KEYs. The KEY corresponds
%% to the KEY in the \bibitem in the reference list below. 

\section{Introduction}

Identifying the sources responsible for the epoch of reionization
(EoR) is still an open and very debated problem in the comprehension
of the various transition phases of the Universe
\citep{meiksin09,mcquinn16,dayal18,giallongo19,wise19}, as witnessed
e.g. by the wide interest on this subject in the last
Decadal Survey on Astronomy and Astrophysics 2020
\citep{alvarez19,cooray19,finkelstein19b,furlanetto19,mould19,papovich19}.

It is now well established that the end of reionization happened relatively
late, rapidly, and through a patchy and strongly inhomogeneous
process. This picture has been validated by the low optical depth
($\tau=0.054$) due to Thomson scattering of cosmic microwave background
(CMB) photons by
\citet{planck18}, indicating a midpoint of reionization at $z\sim
7.8\pm 0.7$ and a duration $\Delta z_{re}\le 2.8$. Recent results by
\citet{planck19} are further lowering down the CMB optical depth,
$\tau=0.0506\pm0.0086$, moving the reionization epoch even later at
$z\sim 7$ \citep[see also]{efstathiou19} and in a shorter time interval
$\Delta z_{re}=1.0^{+1.6}_{-0.7}$ at 68\% confidence level \citep{reichardt20}.

Recently, \citet{mason19} combined constraints from CMB optical depth,
dark gap statistics, Lyman-$\alpha$ damping wing in quasi-stellar objects
(QSOs), and the ratio of Lyman-$\alpha$ emitters (LAEs) over Lyman break
galaxies (LBGs) at $z>6$ to infer an even later mid-point of
reionization, at $z=6.93\pm0.14$. This is also consistent with the
strong and rapid decrease of the photo-ionization rate
$\Gamma_\mathrm{HI}$ observed at $z\ge 5.5$ by
\citet{calverley11,davies18,daloisio18} and by the strengthening and
hardening of the ionizing background at $z\le 5.7$ found by
\citet{becker19}.

The presence of particularly long and deep absorbed troughs in the
spectra of $z\sim 6$ QSOs
\citep{becker15,becker18,keating19,kashino19} marks the position of
the islands of neutral hydrogen, consistent with a very late end of
the reionization process ($z\sim 5.2-5.5$). The analysis of the
probability distribution function of the inter-galactic medium (IGM)
opacity in the Lyman forest of high-z bright quasars
\citep{bosman18,eilers18,kulkarni19,eilers19} confirms that strong
spatial inhomogeneities of neutral hydrogen at $z\lesssim 6$ are
required in order to match these observations, plausibly powered by
the shot noise of rare QSOs \citep{meiksin20}.

The first sources of HI ionizing photons, plausibly star-forming
galaxies (SFGs) and AGNs, were able to close the so-called ``Dark Ages'' at
redshift $z\sim 6-7$, cleaning the fog by neutral hydrogen and
producing the widespread metagalactic ionizing background.
In the last 15 years, the mainstream on reionization was focused
almost {\em totally} on high-z SFGs, with the
exception of \citet{giallongo12,giallongo15,madau15,boutsia18,grazian18,
giallongo19,romano19}. This choice has been driven by four facts:
early WMAP results, heavily affected by dust polarization of the Milky
Way, indicated a much earlier reionization epoch, around $z\ge 12$
\citep{wmap}; the luminosity function of galaxies is gradually
steepening with redshifts at $z\ge 4$ \citep{finkelstein15}; numerous
galaxies exist even at very faint absolute magnitudes ($M_{UV}\sim
-13$) at $z\ge 6$ \citep{livermore16}; at high redshifts, galaxies can
be efficient producers of ionizing photons \citep{bouwens16} and there
were frequent indications of a large emissivity of ionizing photons from
SFGs at $z\ge 3$ \citep[e.g.]{steidel01,shapley06,nestor11,mostardi13}.

Great uncertainties however affect the determination of the Lyman
continuum escape fraction of high-z SFGs. Recent
observations of LBGs at z$\sim$3 by
\citet{steidel18} indicate that the ionizing emissivity from
SFGs, with $f_{esc}^{abs}=9\pm 1$\%, exceeded
that of QSOs by a factor $\sim 3$. Similar results
are obtained recently by \citet{iwata19}, who infer a 3$\sigma$ upper
limits of $f_{esc}^{abs}<8$\% for galaxies at z=3.1 with absolute UV
magnitude $M_{UV}<-18.8$ ($f_{esc}^{abs}<6.3$\% if only IGM attenuation
is taken into account). Similarly, \citet{bian20} derived a stringent
upper limit of 14-32\% at 3$\sigma$ to the Lyman continuum
escape fraction of faint ($M_{1500}\sim -18.8$)
LAEs ($EW\sim 140$ \AA) at $z\sim 3.1$, confirming previous
results by \citet{grazian16,grazian17}, while \citet{bosman20}
reported the measurement of $f_{esc}^{abs}<1$\% for a double peak
LAE at $z\sim 5.7$
in the proximity zone of a bright QSO in the background.
\citet{jones18} combined deep UV imaging
with HST and deep spectra by Keck/DEIMOS in the GOODS-North field to
search for candidate ionizing sources at $z\sim 2.5-3$. At variance
with \citet{steidel18} and \citet{iwata19}, they find that four
candidates out of six are contaminated by foreground galaxies at
lower-z, the ionizing flux in the GOODS-North region is dominated by a
$z\sim 2.6$ AGN and the remaining candidate galaxy has low escape
fraction, well below the required ionization level of the IGM at these
redshifts. AGNs could thus provide a significant contribution to
the ionizing background
at these redshifts. A low Lyman continuum escape fraction $f_{esc}^{abs}<1.5$\%
(at 98\% c.l.) has been derived by \cite{tanvir18} for the host
galaxies of $1.6\le z\le 6.7$ long-duration gamma-ray bursts (GRBs),
assuming they trace
the locations of the massive stars dominating ionizing photon
production. Interestingly, they do not find correlation of the escape
fraction with the galaxy UV luminosity or host stellar mass,
suggesting that faint galaxies at high-z should not dominate the
photon budget required for HI reionization.

Reionization models where ultra-faint dwarf SFGs are
the main contributors to the UV background \citep[e.g.]{finkelstein19} seem
also disfavored by the recent scenarios of late and quick reionization
mentioned above \citep{mason19,naidu19}, since faint dwarf galaxies
would start the reionization process too early, in tension with the
recent results by Planck and with the evolution of the neutral
hydrogen fraction with redshift $x_{HI}(z)$ \citep{hoag19,keating19,yung20}.

It is interesting at this regard the fact that the sources with
suspected or confirmed Lyman continuum radiation at high redshifts are peculiar
and rare, bright SFGs with rather hard ionizing
spectra marked by high ionization emission lines (e.g. NV,
CIV, HeII, OIII, CIII). Their
presence at high redshift can be hardly explained by pure SFGs,
even requiring uncommon assumptions (e.g. large stellar
rotation, binary stellar population, top heavy initial mass function (IMF),
extremely low metallicity) as discussed by a number of recent works
\citep[see]{bradley14,bowler14,kehrig15,stark15,stark16,jaskot16,
senchyna17,nakajima18,berg18,chisholm19,jaskot19,lefevre19,
nanayakkara19,senchyna19a,senchyna19b,schaerer19,stanway19}.
The majority of galaxies showing Lyman continuum emission (both at
low-z and at $z\sim 3-4$) populate the upper end of the BPT diagram \citep{bpt}
or occupy a region of the high ionization line ratio in between SFGs and
AGNs or mainly populated by AGNs (see e.g. Fig. 11 and 14 by Nakajima
et al. 2018). Indeed, \cite{lefevre19} find a marginal 2$\sigma$
detection in the X-ray stacking of strong CIII emitters at
$2<z<4$, consistent with the presence of low-luminosity AGNs.
Interestingly, it is recently emerging the evidence that local
confirmed Lyman continuum emitters or low-z Green Peas, blue compact galaxies,
Lyman break analogs, that are usually
associated with reliable Lyman continuum candidates, are characterized by
significant X-ray emission, not compatible with star-formation
activity but more plausibly powered by low-luminosity AGN activity or
by a large population of high-mass X-ray binaries
\citep{kaaret17,svoboda18,baldassarre19,bao19,bluem19,latimer19,
plat19,prescott19,senchyna19a,senchyna19b,wu19,birchall20,dittenber20}.
It is thus possible that pure stellar radiation from
SFGs is a negligible source of HI ionizing radiation, and
the bulk of Lyman continuum photons escaping at low and high-z are produced
instead by accretion onto super massive black holes (SMBHs).

Recently, a revival of the role of bright QSOs and faint AGNs in the
EoR is progressively emerging. \citet{giallongo15,giallongo19} found
faint AGNs at $z>4$ in the CANDELS fields using the deep Chandra X-ray
imaging available in these fields. Sources were selected in the
HST H-band as having $H<27$ and $z>4$ (which corresponds to an UV
rest-frame selection). AGN candidates were selected looking for
significant X-ray emission in the H-band position of sources with
photometric or spectroscopic redshifts greater than 4. The analysis
suggested a dominant contribution of AGNs to the expected UV background up to
$z\sim 5$ and a possible important contribution up to $z\sim 6$
depending on the uncertainties in the evolution of the AGN luminosity
function at $z>5$ and on the adopted average escape fraction and mean
free path of ionizing photons into the IGM.

Both \citet{cristiani16} and \citet{grazian18} indicated
high escape fractions ($\sim75-80$\%) for QSOs and AGNs
at $z\sim 4$ in a wide luminosity range
($M_{1450}=-30\div -23$), while \citet{romano19} showed that, at
$z\sim 3.6-4.6$, the ionizing background produced by QSOs should be
corrected upward by factor $\sim$1.2-1.7 with respect to previous
estimates in the literature, due to the longer mean free path of HI
ionizing photons. If these values and trends are extrapolated to
higher redshifts, then the AGN population can play a significant role
in the cosmic reionization. At present, however, a notable source of
uncertainty in this respect is the value of Lyman
continuum escape fraction of faint type 2 AGNs at high-z, which could be
significantly low \citep[e.g.]{cowie09,micheva17}.

The main criticism\footnote{The issues of a too early HeII reionization
and the high IGM temperature caused by the hard AGN radiation will
be discussed in Section 5.} against an HI reionization driven by accreting
SMBHs is the dramatically rapid drop of
the space density of AGNs of intermediate luminosity at $z\ge 4$, as
resulting from various color-selected surveys \citep{fan01,cowie09,
parsa18,akiyama18,kim18,matsuoka18,yang18,kulkarni18,cowie20}.
These results, however, could be affected by incompleteness in the
observed number of high-z QSOs. For example, at $z\sim 4$ an almost
spectroscopically ``complete'' sample of $M_{1450}\sim -24$ AGNs by
\citet{boutsia18} based on a multi-wavelength selection which includes
X-ray detections, highlights the possible under-estimation provided by
previous analysis \citep[e.g.]{akiyama18,parsa18}, mainly based on standard
color selections.
Also at very bright magnitudes
($M_{1450}\le -27$), \cite{schindler18,schindler19} find an increase
of $\sim$36\% to the number density of QSOs at $3<z<5$ with respect to
the previous results based on SDSS survey \citep{fan01}. Moreover,
they suggest a moderate evolution of the QSO number density with
redshift and a steeper bright-end luminosity function compared to the
previous SDSS results \citep{fan01}. 

The \citet{giallongo19} luminosity functions are based on AGN
candidates, photometric redshift selections, uncertain corrections due
to the X-ray vs optical flux ratio distribution, and thus they are
still uncertain, especially at $z>5$ and $M_{1450}\sim -23$, where
most of the AGN contribution to the UV background is expected.
We provide here an updated estimate of the AGN space density in the
rest frame UV at $z>5$ and $M_{1450}\sim -22.5$, thanks to the new
analysis of two spectroscopically confirmed AGNs at $z>5$ which have
been optically selected in the CANDELS/GOODS-North and South fields.

The paper is organized as follows: in Section \ref{sec:data} we
describe the data used in this work, in Section 3 we outline the
adopted method to measure the space density of AGNs at $z>5$, and in
Section 4 we show the results on the AGN luminosity function and HI
photo-ionization rate. Section
5 discusses the reliability of our results and conclusions are
provided in Section 6. Throughout the paper, we assume the $\Lambda$
cold dark matter ($\Lambda$-CDM) concordance cosmological model, with
round values H$_0=70$ km/s/Mpc, $\Omega_m=0.3$, and $\Omega_\Lambda=0.7$.
Apparent magnitudes are in the AB photometric system.

%--------------------------------------------------------------------

\section{Data}\label{sec:data}

Two relatively faint AGNs have been spectroscopically confirmed at
$z_{spec}>5$ in the CANDELS footprint, i.e. GDN3333 in the GOODS-North
field and GDS3073 in GOODS-South. We describe here the dataset
available for these two AGNs.

\subsection{GDN3333}\label{sec:gdn3333}

The AGN GDN3333 lies in the CANDELS/GOODS-North field, at
coordinates RA=12:36:47.96 DEC=+62:09:41.3 (J2000) and with an observed
magnitude of 23.91 in the F850LP filter of HST. It has been detected
in X-ray (0.5-8.0 keV) by the ultradeep 2 Msec Chandra X-ray image of
the Chandra Deep Field North (CDFN), as discussed in
\citet{alexander03}. It corresponds to the source with ID=247 in the catalog by
\citet{alexander03}, and it has 138 counts in the 0.5-8.0 keV
band. This source has been selected also in the AGN sample by
\citet{giallongo19} with a flux of 27.2$\times 10^{-17}erg s^{-1}$ in
the 0.5-2.0 keV band. It is clearly detected in X-ray, with a low
probability of spurious detection of $<0.1 \times 10^{-5}$.

Spectroscopic information for GDN3333 has been collected in
\citet{barger02}. GDN3333 has a spectroscopic redshift of $z=5.186$
and, in their catalog, it corresponds to the object with
ID=174 (see their Fig. 6). It shows a strong
Lyman-$\alpha$ line in emission, which is relatively narrow, with a
weak NV 1240 line, also in emission. The Lyman-$\beta$ line is barely
visible. No other emission lines are seen at
$1200\le \lambda_{rest}\le 1600$ {\AA}, since the optical spectrum of GDN3333
is relatively noisy due to strong sky emissions. Based on the available UV
rest-frame spectrum only, this object would have been classified
as a simple SFG, if it had not been detected in the
deep Chandra image in X-ray by \citet{alexander03} or
by \citet{giallongo19}. It is worth noting that the NV 1240 line falls
on a strong night-sky emission line, thus it is not obvious to assure
its presence in the available spectrum by \citet{barger02}.

Summarizing, given its clear X-ray detection by Chandra, as shown in
\citet{alexander03} and its secure spectroscopic redshift at z=5.186
provided by \citet{barger02}, there is no doubt that GDN3333 is an
AGN, as concluded also by \citet{giallongo19}. Moreover, its X-ray
over optical flux ratio is $\sim 0.1$ (see Fig. 2 of Giallongo et
al. 2019), which is one of the highest X/optical flux ratio among the
AGN candidates at magnitudes $H_{160}\le 24$ provided by
\citet{giallongo19}. Based on these considerations, we conclude that
its X-ray flux cannot come from star formation activity, but it is
powered by an accreting SMBH. GDN3333 is the only known AGN
brighter than $H=24$ at $z>5$ in the CANDELS fields and it has already
been included in the AGN luminosity function at $z\sim 5.5$ by
\citet{giallongo19}.

\subsection{GDS3073}\label{sec:gds3073}

The source GDS3073 has sky coordinate RA=03:32:18.92 DEC=-27:53:02.7
and belongs to the CANDELS/GOODS-South field. Its F850LP magnitude is
24.52 and its spectroscopic redshift is $z\sim 5.56$
\citep{raiter10,vanzella10}. GDS3073 has a spatially compact
morphology in the ACS HST images and resemble an almost unresolved
point-like source, similar to GDN3333. This object is not detected in
the ultradeep 7 Msec X-ray image by Chandra down to a flux limit of
$10^{-17}erg s^{-1}$ in the 0.5-2.0 keV band \citep{giallongo19}.
GDS3073 has been detected by ALMA in CII at 158$\,\mu$m
by the ALPINE survey (``ALMA
Large Program to INvestigate C+ at Early Times''), as described in
\citet{alpine}. In the following, we discuss three deep
spectroscopic observations of GDS3073 with FORS2, X-Shooter, and VIMOS
at the VLT telescope. These spectra indicate that GDS3073 is a
bona-fide AGN, as we show in the following.
At the end of this section we also summarize the sub-mm properties of
GDS3073 by the ALPINE survey with ALMA.

\subsubsection{The FORS2 data}\label{sub:fors}

GDS3073 has been selected for the FORS2 spectroscopic observations as
a F606W-band dropout and has a blue F775W-F850LP color due to a
strong Lyman-$\alpha$ emission line falling in the F775W band
\citep{raiter10}. The spectroscopic observations of GDS3073 have been
carried out with the ESO VLT FORS2 instrument, yielding a wavelength
coverage of approximately 0.55-1 $\mu$m with a resolving power of
$R\sim 660$. The total exposure time on GDS3073 is of 4 hours
\citep{vanzella10}.

\begin{figure}
\centering
\includegraphics[width=7cm,angle=-90]{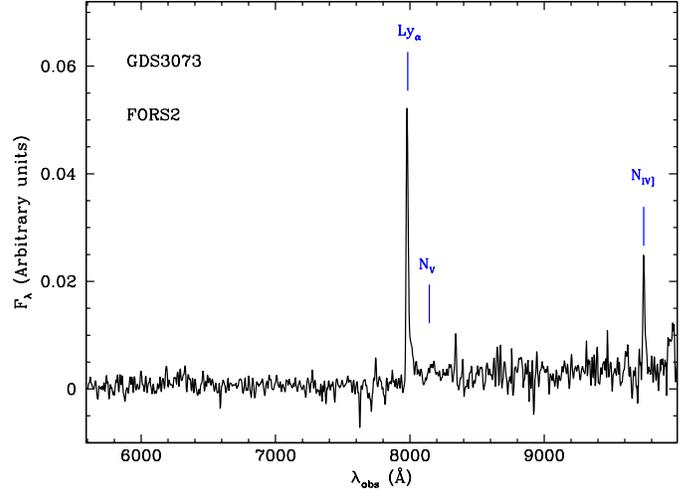}
\caption{The FORS2 spectrum of GDS3073.
The expected positions of common emission lines for galaxies and AGNs are
highlighted by vertical blue segments. The NV 1240 emission
line at $\lambda_{obs}\sim 8140$ {\AA}, corresponding to $z_{spec}=5.563$,
is barely visible in the FORS2 spectrum.
}
\label{fig:specfors}
\end{figure}

The FORS2 spectrum of GDS3073 is shown in Fig. 4 of \citet{raiter10},
in Fig. 1 of \citet{vanzella10} and in Fig. \ref{fig:specfors} of this
paper. GDS3073 shows a strong and asymmetric Lyman-$\alpha$ line, with
rest-frame equivalent width (EW) of $\sim 60$ {\AA}
and a peak at $\lambda_{obs}=7978$ {\AA},
corresponding to a redshift of $z=5.563$. It also shows an emission
line at the observed wavelength of 9735 {\AA}. This line has been
associated to the inter-combination doublet of NIV] at 1483,
1486 {\AA} rest-frame at a redshift of $z_{spec}=5.563$, in
agreement with the Lyman-$\alpha$ line in emission.
The presence of significant NIV] in emission is rather unusual
in known astronomical objects, but it is somehow detected in rare
``Nitrogen Loud'' QSOs \citep{baldwin03,glikman07},
where they usually show weak or absent CIV 1549 in emission.
Such features have been also detected in the Lynx arc at z=3.4
\citep{fosbury03}.

From a detailed fitting of the spectral energy distribution of
GDS3073, extending to the IRAC bands, there are clear indications of
the presence of strong nebular emission of OIII 5007 and H-$\beta$
\citep{raiter10,vanzella10}. Combined with its compact morphology,
GDS3073 could be classified as a Green Pea analog at high-z
\citep{cardamone09}.

The FORS2 spectrum of GDS3073 (Fig.\ref{fig:specfors}) shows also the
possible presence of NV 1240 emission line at $z_{spec}=5.563$
($\lambda_{obs}\sim 8140$ {\AA}), although at low
significance. This spectrum is not deep enough for a clear detection,
but already at this stage the NV line emerges from the
continuum. From the FORS2 spectrum alone, it is not possible to
conclude whether GDS3073 is a SFG or an AGN, therefore we
analyze also the VIMOS and X-Shooter spectra to have further
information about its nature.

\subsubsection{The VIMOS data}\label{sub:vimos}

GDS3073 has been observed for 20 hours with the VIMOS spectrograph
during the ESO program ID 194.A-2003 \citep{mclure18}. The raw data from
the ESO Science Archive have been reduced with a custom pipeline,
described in detail in \citet{grazian18}.

\begin{figure}
\centering
\includegraphics[width=7cm,angle=-90]{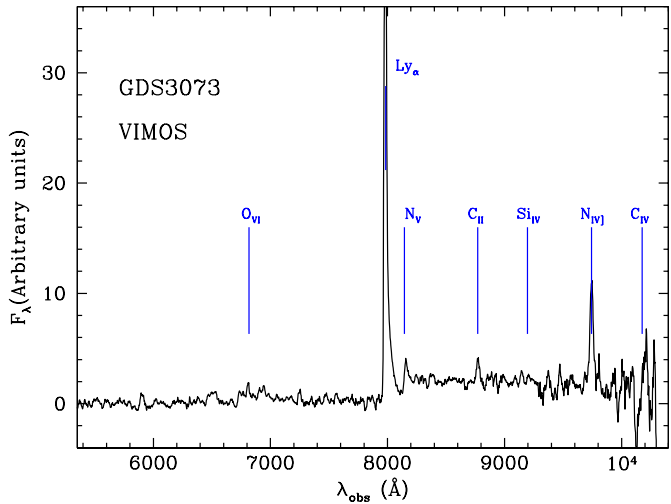}
\caption{The VIMOS 1-dimensional spectrum of GDS3073, where emission
lines typical of AGNs are highlighted by the blue vertical segments. In
particular, the NV 1240 and OVI 1032-1038 emission lines
unambiguously mark the AGN nature of this object.
}
\label{fig:specvimos}
\end{figure}

Fig.\ref{fig:specvimos} shows the 1-dimensional VIMOS deep spectrum of
object GDS3073. Strong emission lines of high ionization (NV
1240, NIV] 1483-1486) are clearly detected in the spectrum, while
other high ionization lines (Lyman-$\beta$, OVI 1032-1038) are
barely visible. The CIV line at 1549.1 {\AA} rest-frame is
probably present, but it falls on a sky emission line and it is only
barely visible after 20 hours of exposure. In particular, the
NV 1240 and OVI 1032-1038 emission lines unambiguously
ensure the AGN nature of this object. It is worth noting that the
NV 1240 emission line was not clearly detected in the FORS2
spectrum by \citet{raiter10,vanzella10}, after 4 hours of exposure
time, while it is clearly detected here after 20 hours of VIMOS. It
clearly indicates that deep exposures are required in order to
determine the true nature (SFG vs AGN) of high redshift objects.

\begin{figure}
\centering
\includegraphics[width=9cm,angle=0]{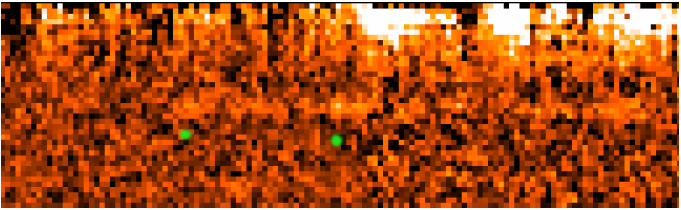}
\caption{A zoom to the VIMOS 2-dimensional spectrum of GDS3073 close to
$\lambda_{obs}\sim 6800$ {\AA}. The two green spots identify the wavelength
position of the emission lines Lyman-$\beta$ (left) and OVI
1032-1038 (right)
at $z_{spec}=5.563$. These emission lines are typical of AGNs. The bright
white spots in the upper side of the spectrum are the continuum of a bright
object which is close to the GDS3073 source.
}
\label{fig:specvimos2d}
\end{figure}

Fig.\ref{fig:specvimos2d} shows a zoom of the VIMOS GDS3073
2-dimensional spectrum close to the $\lambda_{obs}\sim 6800$ {\AA}
wavelength range. The OVI 1032-1038 emission line is clearly
visible in the spectrum. The OVI 1032-1038 line is clearly detected
at a S/N ratio of 6.7. This line unambiguously confirms the AGN
nature for GDS3073. Moreover, the Lyman-$\beta$ line in emission is
barely detected (S/N=4.4). Also this line is very rare in
SFGs, while it is quite common in AGNs and QSOs.

The flux ratio between the Lyman-$\alpha$ line and the NV 1240 in the
VIMOS spectrum of GDS3073 is 6.6, which is significantly lower than
the ratio measured in LBGs by
\citet{shapley03} (Ly-$\alpha$/NV=15) at $z\sim 3$ or at higher redshifts. This
ratio is lower than or more similar to the Ly-$\alpha$/NV ratio in faint
high-z AGNs both at $z<3$ \citep{steidel02,hainline11} and at
$z>6$ \citep{hu17,sobral17,laporte17,mainali18}, with Ly-$\alpha$/NV=1-9.
This measurement further corroborates the AGN nature of GDS3073.

\subsubsection{The X-Shooter data}\label{sub:shooter}

\begin{figure}
\centering
\includegraphics[width=9cm,angle=0]{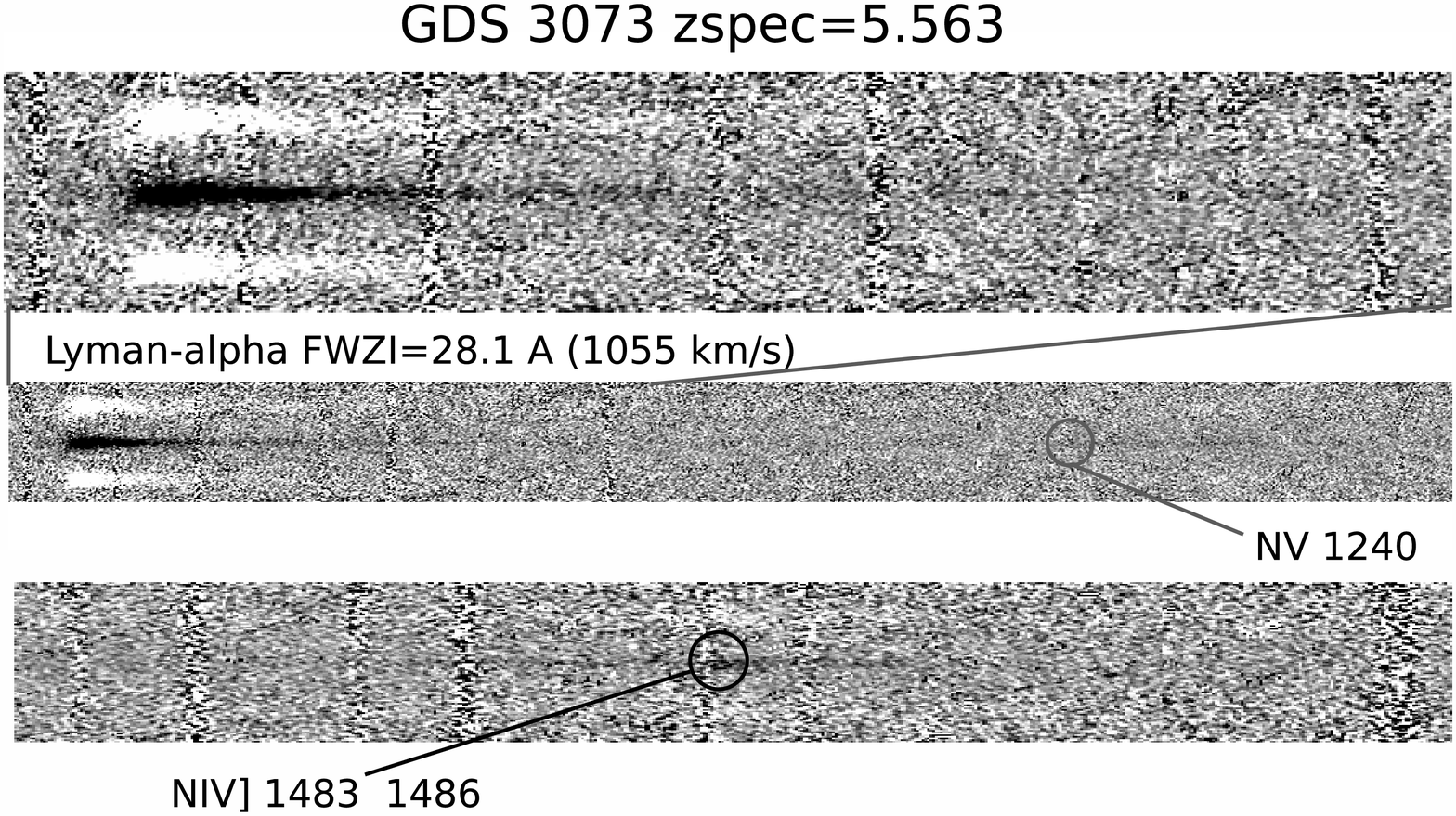}
\includegraphics[width=9cm,angle=0]{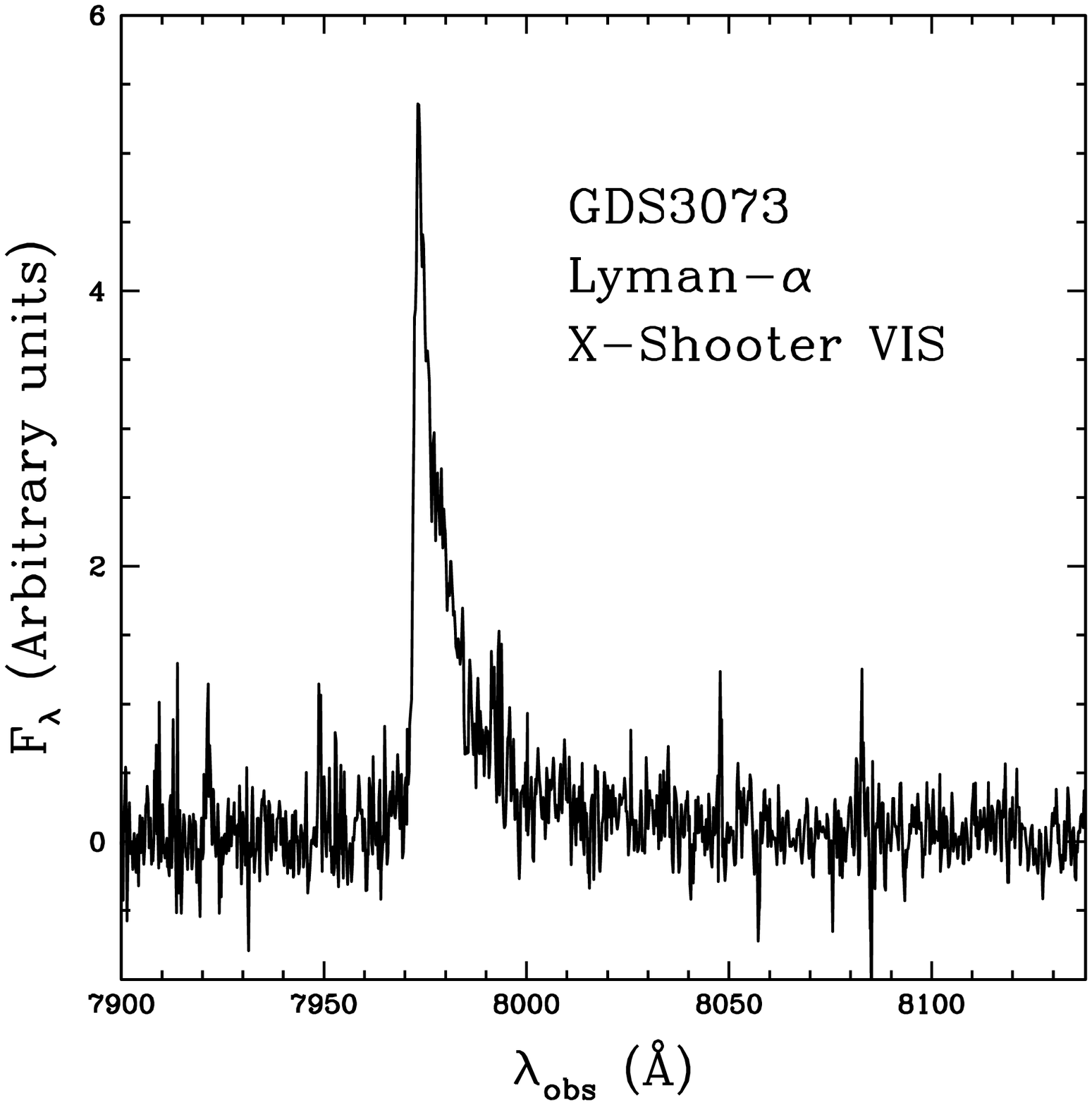}
\caption{{\em Top}: A zoom to the X-Shooter 2-dimensional spectrum
of GDS3073 showing the Lyman-$\alpha$, NV 1240, and NIV]
1483-1486 emission lines at $z_{spec}=5.563$. {\em Bottom}: The Lyman-$\alpha$
emission line observed by X-Shooter shows a full width at zero
intensity (FWZI) of at least 28 {\AA}, corresponding to more than
1000 km/s velocity. These outflow velocities are typical of AGNs.
}
\label{fig:xshooter}
\end{figure}

GDS3073 has been observed by the X-Shooter spectrograph under two
observing programs (384.A-0886 in ESO period 84 and 089.A-0679 in ESO
period 89) for a total of 49 hours of net exposure time (21 hours in
P84 and 28 hours in P89). The raw data have been reduced with the
X-Shooter pipeline available under the ESO REFLEX Pipeline
environment\footnote{https://www.eso.org/sci/software/esoreflex/}. We
stack together all the 49 hours of exposure time in a single image, in
order to maximize the signal to noise ratio of the resulting spectrum.
The X-Shooter spectrum of GDS3073 is rather noisy and only the
Lyman-$\alpha$ and NIV] 1483-1486 emission lines are clearly
detected (Fig. \ref{fig:xshooter}, top). There is a hint of
OVI 1032-1038 and NV 1240 emission lines, but the S/N
ratios of these transitions are quite low due to the higher spectral
resolution w.r.t. the VIMOS spectrum, despite the longer exposure time
of the X-Shooter spectrum.  The CIV 1549 emission line is
barely visible in the X-Shooter spectrum, but it falls at the limit of
the VIS arm ($\lambda_{obs}\sim 10200$ {\AA}) and its S/N ratio is
very low. Unfortunately, due to the much higher spectral resolution,
the X-Shooter spectrum does not add too much information with respect
to the VIMOS and FORS2 spectra, but it is useful to confirm
independently the presence of OVI, NV, and possibly
CIV emission lines. Thus the FORS2, VIMOS, and X-Shooter deep
spectra together confirm that GDS3073 is clearly an AGN with narrow
emission lines.

Fig. \ref{fig:xshooter} illustrates a small portion of the
2-dimensional spectrum of GDS3073, showing the details of the
Lyman-$\alpha$ emission line observed by X-Shooter. Its large
extension, with a full width at zero intensity (FWZI) of at least 28
{\AA}, corresponds to a velocity of 1055 km/s. Such
velocities are typical of the outflows expected in faint AGNs in the
primeval universe \citep{menci19}.

\subsubsection{ALMA data}\label{sub:alma}

In the ALPINE survey, GDS3073 is labeled CANDELS-GOODSS-14 and
has a systemic redshift of $z=5.5527$ based on the CII 157.7 $\mu$m line
\citep{alpine,bethermin20,cassata20}.
In the ALMA datacube, GDS3073 shows a compact
morphology (unresolved at the ALMA spatial resolution of 0.7 arcsec
beam), and its kinematics are mainly dominated by velocity dispersion.
The large redshift difference between the Lyman-$\alpha$ and the C+ emissions
corresponds to $\sim 500$ km/s \citep{cassata20}.
In the ALMA C+ map of GDS3073, a
faint tail is present, extending towards the north-west of the optical
source, as shown in Fig. 4 of \citet{alpine}. The large spatial and
velocity offset, which is an almost frequent case for ALMA detected
sources \citep[e.g.]{carniani17}, could be a plausible indication
of an on-going merger of GDS3073 with a dusty companion galaxy. This
merger could be at the origin of the AGN activity for GDS3073.

%--------------------------------------------------------------------

\section{The Method}\label{sec:method}

The availability of two faint ($L\sim 0.1 L^*$) AGNs, spectroscopically
confirmed at $z>5$ in a relatively small area covered by the
CANDELS-GOODS survey, allows an improved estimate of the space density
of high-z accreting SMBHs. The computation of the AGN Luminosity
Function at $z\sim 5.5$ in the CANDELS fields has been carried out
following the same technique adopted in \citet{boutsia18}. In order to
derive an estimate of the space density of faint AGNs,
we decide to use in the computation only the spectroscopically
confirmed $z>5$ AGNs analyzed here, i.e. GDN3333 and GDS3073. We start
from their $Z_{850}$ HST magnitudes, taken from the CANDELS official
catalogs \citep{guo13,barro19}, and from their spectroscopic
redshifts, which have been summarized here in Table \ref{tab:sample}.

We derive the absolute magnitude at 1450 {\AA} rest frame (M$_{1450}$)
for each source from the $Z_{850}$ HST magnitude and by applying a
K-correction according to the following equation:
\begin{equation}
M_{1450} = Z_{850}-5log(d_L)+5-2.5log(1+z_{spec})+K_{corr}
\end{equation}
where $d_L$ is the luminosity distance, expressed in pc, and the
k-correction is given by
\begin{equation}
K_{corr}=2.5\alpha_\nu log_{10}(\lambda_{obs}/(1+z_{spec})/\lambda_{rest}) \, .
\end{equation}

The AGN intrinsic slope $\alpha_\nu$ has been fixed to -0.7, while
$\lambda_{obs}=8879$ {\AA} is the central wavelength of the $Z_{850}$
filter, and $\lambda_{rest}=1450$ {\AA}. The choice of the $Z_{850}$
band for M$_{1450}$ calculation allows us to minimize the k-correction.
The absolute magnitudes M$_{1450}$ have been included in
Table \ref{tab:sample}.

\begin{table}
\caption{The $z_{spec}>5$ AGNs in the CANDELS/GOODS fields.}
\label{tab:sample}
\begin{center}
\begin{tabular}{c c c c}
\hline
\hline
Name & z$_{spec}$ & $Z_{850}$ & M$_{1450}$ \\
\hline
GDN3333 & 5.186 & 23.91 & -22.56 \\
GDS3073 & 5.563 & 24.52 & -22.10 \\
\hline
\hline
\end{tabular}
\end{center}
\end{table}

The AGN space density at $M_{1450}=-22.5$ has been computed by adopting
the $1/V_{max}$ approach, without any correction for incompleteness. We
want indeed to be sure that our luminosity function is a reliable
estimate and it is not plagued by possible uncertainties. In order to
compare our results with the recent ones derived in the CANDELS fields
by \citet{giallongo19}, we adopt the same area and the same redshift
interval. The total area used here is 551.5 sq. arcmin., which is the
sum of the individual HST pointings of the CANDELS/GOODS-North,
GOODS-South, and EGS. The adopted redshift interval is $5.0<z<6.1$. It
is worth noting that in the EGS field the spectroscopic search for
high-z objects has not been carried out extensively as in the GOODS
fields, so it is possible that other $z>5$ AGNs are present in this
field. In principle, we should exclude the EGS field from the volume
computation here. We decided however to take into account also the EGS
fields in order to be comparable with the results of
\citet{giallongo19}. Of course, if we reduce the area only to the two
GOODS fields, for a total of 346 sq. arcmin., then the AGN space density
provided here should be increased by a factor of $\sim 1.6$.

%-----------------------------------------------------------------

\section{Results}\label{sec:results}

\subsection{The luminosity Function
of faint AGNs at $z\sim 5.5$}\label{sub:lfz5}

The space density of faint ($L\sim 0.1L^*$, M$_{1450}\sim -22.5$) AGNs
at $z\sim 5.5$ has been derived in the CANDELS fields as described in
Section \ref{sec:method}. It is worth noting here that the CANDELS
luminosity function has been derived by simply dividing the actual
number of the spectroscopically confirmed AGNs by the comoving volume
between $5.0\leq z\leq 6.1$. No correction for any kind of
incompleteness has been applied to the measured density, and
any incompleteness corrections would move the space
density of $z\sim 5.5$ AGNs upwards.
There could be indeed other faint high-z
AGNs which have not yet spectroscopically identified or with shallow
spectra that do not allow a reliable classification, as happened in the past
with old data for GDN3333 and GDS3073.

\begin{table}
\caption{The space density $\Phi$ of $z\sim 5.5$ AGNs
in the CANDELS fields.}
\label{tab:lfobs}
\begin{center}
\begin{tabular}{c c c c c}
\hline
\hline
z$_{spec}$ & M$_{1450}$ & $\Phi$ & $\sigma_\Phi(up)$ & $\sigma_\Phi(low)$ \\
 & & $10^{-6} cMpc^{-3}$ & $10^{-6} cMpc^{-3}$ & $10^{-6} cMpc^{-3}$ \\
\hline
5.5 & -22.33 & 1.291 & 1.717 & 0.854 \\
\hline
\hline
\end{tabular}
\end{center}
\end{table}

\begin{figure}
\centering
\includegraphics[width=9cm,angle=0]{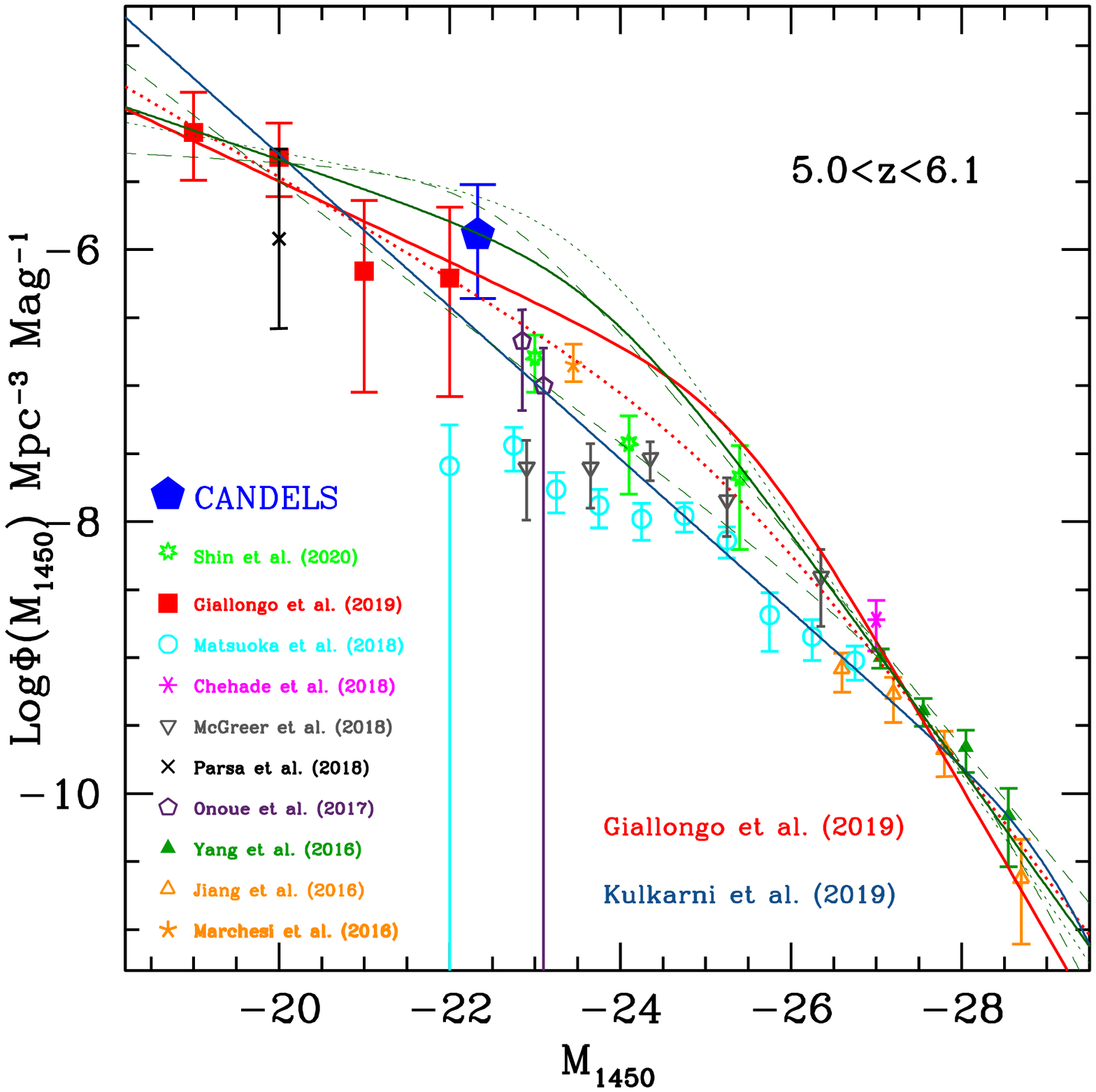}
\caption{The luminosity function of AGNs at $z\sim 5.5$.
The filled blue pentagon shows the
space density obtained from the CANDELS fields (with objects GDN3333 and
GDS3073). The continuous dark green curve is the best fit provided in this
work, the dashed dark green lines indicate the region corresponding to
$\pm 1\sigma$ uncertainty in the luminosity function parameters, while
the dotted dark green line mark the luminosity function with the maximum
ionizing background allowed by the data (see text for details).
The red filled squares are the luminosity function recently
derived by \citet{giallongo19}, while the dotted and continuous red lines are
their best fit luminosity functions (model 3 and 4, respectively).
The blue continuous curve is the best fit AGN luminosity function by
\citet{kulkarni19}.
The other open symbols are a summary of recent luminosity
function determinations at $z\sim 5-6$. The point from \citet{marchesi16}
has been converted into UV rest frame by \citet{ricci17} based on a
standard X-ray to UV luminosity ratio distribution.
All the data points and curves
have been shifted to $z=5.5$ adopting the density evolution recipe
by \citet{fan01}.
}
\label{fig:lfz5}
\end{figure}

Table \ref{tab:lfobs} reports the space density of $z\sim 5.5$ AGNs
in the CANDELS fields.
Fig.\ref{fig:lfz5} summarizes the resulting luminosity function of
QSOs/AGNs at $z\sim 5.5$. It shows the space density obtained from
the CANDELS fields (filled blue pentagon), using objects GDN3333 and
GDS3073. Error bars to the space density have been derived by adopting the
statistics of \citet{gehrels86} for low number counts.
This plot compares the luminosity function obtained in this
paper with the one (red filled squares) recently derived by
\citet{giallongo19}. Other small open symbols in Fig.\ref{fig:lfz5}
are a summary of the AGN space density determinations at $z\sim 5-6$
\citep{jiang16,yang16,marchesi16,ricci17,onoue17,parsa18,mcgreer18,
chehade18,matsuoka18,shin20}.
The dotted and continuous red lines in Fig.\ref{fig:lfz5}
are the \citet{giallongo19} best
fit luminosity functions at $z\sim 5.5$. In particular, the dotted line
is the best fit obtained by leaving all parameters free during the fitting
procedure (their model 3), while the red continuous curve has been
obtained by fixing the two slopes to the best fit values derived at $z=4.5$
(model 4). The blue continuous curve is the best fit AGN luminosity
function by \citet{kulkarni19}. When the mean redshifts of the
various luminosity functions displayed here are not centered at
$z=5.5$, i.e. the mean value of our survey, we correct the original
densities, taken from the literature, by a factor
$\Phi_{corr}=\Phi_{orig}\times 10^{k(z_{orig}-5.5)}$,
with $k=0.47$, as found by \citet{fan01}.
Table \ref{tab:lflit} provides the
complete list of the AGN space densities at $z\sim 5-6$ from recent works,
as shown in Fig.\ref{fig:lfz5}.

\begin{table*}
\caption{The space densities $\Phi$ of QSOs and AGNs at $z\sim 5-6$
from recent works.}
\label{tab:lflit}
\begin{center}
\begin{tabular}{c c c c c c}
\hline
\hline
paper & redshift & M$_{1450}$ & Selection criterion &
$N_{obj}$ & $Area(deg^2)$ \\
\hline
\hline
\multicolumn{6}{c}{Included in the AGN's luminosity function fitting
procedure} \\
\hline
\citet{jiang16} & 5.7-6.4 & -29.0,-27.0 & optical colors & 52 & 11240 \\
\citet{yang16} & 4.7-5.4 & -29.3,-26.8 & optical colors & 99 & 14555 \\
\citet{chehade18} & 5.7-6.4 & -27.5,-26.5 & optical/NIR colors & 6 & 3119 \\
\citet{giallongo19} & 5.0-6.1 & -20.0,-19.0 & NIR/X-ray & 9 & 0.15 \\
\hline
\hline
\multicolumn{6}{c}{Excluded from the AGN's luminosity function fitting
procedure} \\
\hline
\citet{marchesi16} & 3.0-6.85 & $Log(Lx)>43.55$ & X-ray & 174 & 2.2 \\
\citet{ricci17} & 5.0-6.5 & -22.5,-18.5 & X-ray to UV conversion & \nodata & \nodata \\
\citet{onoue17} & 5.7-6.5 & -23.5,-22.5 & optical colors & 2 & 6.5 \\
\citet{parsa18} & 5.0-6.5 & -21.0,-19.0 & NIR/X-ray & 1 & 0.09 \\
\citet{mcgreer18} & 4.7-5.4 & -26.4,-22.4 & optical/NIR colors & 37 & 105 \\
\citet{matsuoka18} & 5.7-6.5 & -26.5,-22.0 & optical colors & 110 & 650 \\
\citet{shin20} & 4.6-5.4 & -26.0,-23.0 & optical/NIR colors & 10 & 6.5 \\
\hline
\hline
\end{tabular}
\end{center}
A collection from the recent literature of the space densities of QSOs
and AGNs at $z\sim 5-6$, shown in Fig.\ref{fig:lfz5}. The upper part
of the table summarizes the space densities used for the luminosity
function fitting, while the lower part of the table includes space
densities, typically associated to faint absolute magnitudes, which
have not been considered during the fitting process.
Only the faintest data points of \citet{giallongo19} have been considered
for our luminosity function derivation. The \citet{giallongo19} luminosity
function is based on a rest-frame UV (observed NIR $H_{160}\le 27$ AB magnitude)
selected sample with photometric redshift greater than 4 and with
robust associations to X-ray sources. The X-ray luminosity function
from \citet{marchesi16} has been converted into an UV luminosity function
by \citet{ricci17} based on a standard X-ray/UV luminosity ratio distribution.
\end{table*}

As can be seen from Figure \ref{fig:lfz5}, the CANDELS space density
(blue filled pentagon), without any corrections, agrees well with and it
is even larger than the luminosity function by \citet{giallongo19} at
$M_{1450}\sim -22$ (red filled squares), which at these luminosities has
one AGN at $z>5$. Our data point is slightly
higher than their fit (model 3, red dotted line), possibly indicating
that the space
density of AGNs is even larger than previous estimates by
\citet{giallongo19}. It is also consistent with their best fit value
(model 4, red continuous line).
At this stage, the AGN space density at $z\sim 5.5$ in CANDELS is
much larger than the luminosity function determinations appeared recently
in the literature. In particular, our luminosity function is ruling
out the results of
\citet{mcgreer18,matsuoka18,cowie20}
and it is only marginally consistent with
\citet{marchesi16,ricci17,onoue17,parsa18,kulkarni19,shen20,shin20}.

Due to the relatively small area of the CANDELS survey, the
cosmological volumes probed by this paper is still limited, and
no AGN brighter than $M_{1450}\sim -23$ has been presently found in
GOODS-North and GOODS-South fields. For this reason, very limited
information exists at $M_{1450}\sim -24.5$ ($L\sim L^*$), i.e. the plausible
position of the break of the AGN luminosity function at $z\sim
5.5$. The recent determinations by \citet{mcgreer18} and
\citet{matsuoka18}, e.g., are probably affected by incompleteness, and they
are possibly underestimating the luminosity function even at the
break. Efficient rest-frame UV color selections coupled with
morphological criteria could severely underestimate the AGN space
density at high redshifts, as found by \citet{boutsia18,stevens18} and
\citet{adams19} at slightly lower redshifts ($z\sim 4$), or at $z\sim 5$
by \citet{shin20}.

We have refined the best fit AGN luminosity function at $z\sim 5.5$
provided in the Equation 2 of \citet{giallongo19} with our new
CANDELS data point at $M_{1450}\sim -22.3$, given in Table
\ref{tab:lfobs}.
We have compared our volume densities derived in the CANDELS fields
with that of the brightest ($M_{1450} \le -27$) QSOs selected in the
SDSS and ATLAS/WISE surveys \citep{jiang16,yang16,chehade18}, where
selection biases with respect to the morphological appearance and
X-ray properties are small (upper part of Table \ref{tab:lflit}). At
these bright magnitudes, indeed, no X-ray selected QSO with strong
absorption in the rest-frame optical/UV is expected, and optical
colors or X-ray selections are both representative of the same global
AGN population.
At fainter magnitudes, incompleteness in surveys based on optical
color becomes significant and the scatter in their observed number
densities is large. For this reason we did not include in our best
fitting procedure the luminosity function points derived from color
selected surveys in the range $-26\lesssim M_{1450}\lesssim -22$ but
we show them in Fig.\ref{fig:lfz5} and in the lower part of Table
\ref{tab:lflit} only for comparison with our results.
The resulting best fit luminosity function after the $\chi^2$
minimization procedure is summarized in Table \ref{tab:lffit} and it is
shown by the continuous green line in Fig.\ref{fig:lfz5}.

\begin{table*}
\caption{The best fit luminosity function of AGNs at $z\sim 5.5$ with the
corresponding HI ionizing emissivity and photo-ionization rate.}
\label{tab:lffit}
\centering
\begin{tabular}{c c c c c c c c}
\hline
\hline
 & z$_{spec}$ & $\Phi^*$ & $M^*$ & $\beta$ & $\gamma$ & $\epsilon^{912}_{24}$
& $\Gamma_{-12}$ \\
 & & $cMpc^{-3}$ & & & & $10^{24}erg/s/Hz/Mpc^3$ & $10^{-12}s^{-1}$ \\
\hline
BEST FIT & 5.5 & $0.78\times 10^{-6}$ & -23.78 & 1.535 & 3.175 & 1.45 & 0.13 \\
\hline
-1$\sigma$ & 5.5 & $0.69\times 10^{-9}$ & -27.68 & 2.205 & 4.055 & 0.16 & 0.02 \\
+1$\sigma$ & 5.5 & $3.89\times 10^{-6}$ & -22.52 & 1.090 & 2.920 & 1.74 & 0.16 \\
\hline
MIN $\Gamma_{-12}$ & 5.5 & $0.69\times 10^{-9}$ & -27.68 & 2.205 & 4.055 & 0.16 & 0.02 \\
MAX $\Gamma_{-12}$ & 5.5 & $2.09\times 10^{-6}$ & -23.54 & 1.305 & 3.320 & 2.42 & 0.22 \\
\hline
\hline
\end{tabular}
\\
$\beta$ and $\gamma$ are the faint and bright slopes, respectively, of
the AGN Luminosity Function at $z\sim 5.5$. The $\pm 1 \sigma$ lines
correspond to the minimum and maximum slopes allowed by the best fit
at 68\% confidence level, respectively. The -1$\sigma$ corresponds to
the minimum photo-ionization rate (MIN $\Gamma_{-12}$) while the
maximum value of the photo-ionization rate (MAX $\Gamma_{-12}$) is not
associated to the +1$\sigma$ limit of the best fit luminosity function
due to the covariance between the four parameters $\Phi^*$, $M^*$,
$\beta$, and $\gamma$.
\end{table*}

The uncertainties on the best fit luminosity function parameters have been
derived by considering all the possible solutions satisfying $\Delta
\chi_{red}^2\le 1.0$, where $\Delta \chi_{red}^2$ is
$(\chi^2-\chi_{min}^2)/ndf$. Here $\chi_{min}^2$ is the minimum value
associated to the best fit solution, while $ndf$ is the number of
degree of freedom, i.e. the number of independent data points used
(eleven) minus the number of fitted parameters, in our case four.
Since we lack reliable space densities around the knee of the luminosity
function, at $M_{1450}\sim -24.5$, the position of the break is presently
unconstrained and the uncertainties span a wide range, from -27.7 to
-22.5, with a best fit at $M^*_{1450}=-23.78$, corresponding to a density
of $Log(\Phi^*)=-6.11$. The lack of suitable data around $L^*$ causes a
strong degeneracy between $M^*_{1450}$ and $Log(\Phi^*)$, with an almost
linear spread from $M^*_{1450}=-27.68$ and $Log(\Phi^*)=-9.16$ up to
$M^*_{1450}=-22.52$ and $Log(\Phi^*)=-5.41$, which corresponds to the 1$\sigma$
limits. The uncertainty on the two slopes of the luminosity function has a
range between 1.1 and 2.2 for the faint-end slope, with a best fit of 1.53,
and an interval of 2.9-4.0 for the bright end slope, with a best fit of 3.2.
The dashed green lines in Fig.\ref{fig:lfz5} correspond to the two
extreme luminosity functions at 1$\sigma$ confidence level.

Following \citet{giallongo19}, we have repeated the best fit of the
luminosity function by fixing the two slopes to $\beta=1.92$ and
$\gamma=3.09$ (model 3) and $\beta=1.74$ and $\gamma=3.72$ (model
4). The position of the knee of the luminosity function is thus at
$M^*_{1450}=-25.06$ and $Log(\Phi^*)=-7.29$ for model 3 and
$M^*_{1450}=-25.37$ and $Log(\Phi^*)=-7.05$ for model 4,
respectively. These values are pretty consistent with the best fit
obtained in \citet{giallongo19}, with a slightly higher normalization.

\subsection{The ionizing background at $z>5$
and implications on Reionization}

\begin{figure}
\centering
\includegraphics[width=9cm,angle=0]{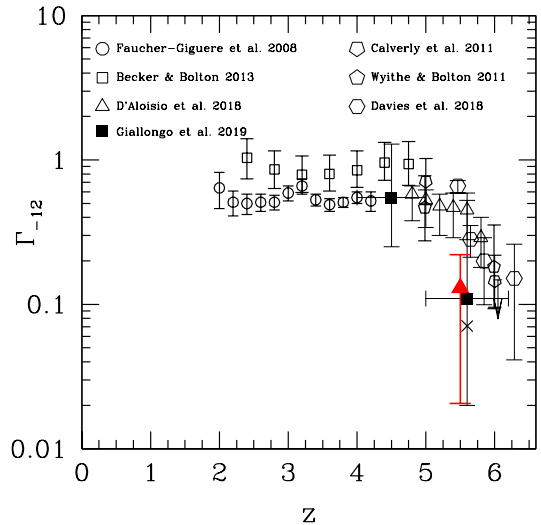}
\caption{
The red filled triangle shows the photo-ionization rate $\Gamma_{-12}$,
in units of $10^{-12} s^{-1}$, produced
by AGNs at $z\sim 5.5$, adopting the luminosity function parameterization
summarized in Table \ref{tab:lffit}. The error bars associated to the red
triangle are the 1$\sigma$ uncertainty in the
photo-ionization rate $\Gamma_{-12}$.
The filled black squares are the
ionizing background of AGNs at $z\sim 4.5$ and $z\sim 5.5$ obtained by
\citet{giallongo19} (their models 2 and 4), while the black cross
at $z\sim 5.5$ is the result of their model 3.
The other open black symbols are the values of $\Gamma_{-12}$
at different redshifts
inferred from the Lyman-$\alpha$ forest analysis in
high-z QSO spectra or through the proximity effect.}
\label{fig:uvb}
\end{figure}

We compute the emissivity $\epsilon^{912}_{24}$ at $\lambda=912$ {\AA}
rest frame by integrating the luminosity function, multiplied by L,
from 0.01$L^*$ to
100$L^*$. An escape fraction of 100\% has been assumed, adopting a
spectral slope of $\alpha_\nu=-0.44$ and of $\alpha_\nu=-1.57$ between
1450 and 1200 {\AA} rest frame and between 1200 and 912 {\AA} rest
frame, respectively \citep{sb03}. This choice
corresponds to a flux ratio of 1.67 between 1450 and 912 {\AA}. The
adopted spectral slopes are similar to the ones derived by
\citet{lusso15}. The Photo-ionization rate
$\Gamma_{-12}=0.13^{+0.09}_{-0.12}$ has been computed following
\citet{giallongo19}, adopting the mean free path of HI ionizing
photons of \citet{worseck14}, which is 9 proper Mpc at $z=5.5$. The
value of $\Gamma_{-12}$ has been increased by a factor of 1.2, in
order to take into account the contribution by radiative recombination
in the IGM \citep{daloisio18}. We leave to Section 4 of
\citet{giallongo19} for further details about the calculations of the
emissivity and photo-ionization rate. In order to derive the 1$\sigma$
interval for $\Gamma_{-12}$ we have considered all the possible
combinations of luminosity function parameters which agree at 68\%
level with the best fit solution, and for each combination of the four
parameters we compute the photo-ionization rate at $z\sim 5.5$. Due to
the degeneracy between $M^*_{1450}$ and $\Phi^*$, outlined in the
previous sub-section, the 1$\sigma$ uncertainty in the
photo-ionization rate $\Gamma_{-12}$ is limited to the interval
0.02-0.22, with a peak probability for $\Gamma_{-12}=0.13$.
This value corresponds to 46\% of the UV background inferred by
\citet{davies18}, or alternatively to the 29\% of the one by \citet{daloisio18}.
The red triangle in Fig.\ref{fig:uvb} shows
the photo-ionization rate $\Gamma_{-12}$,
with its 1$\sigma$ confidence level, produced
by AGNs at $z\sim 5.5$, adopting the luminosity function parameterization
summarized in Table \ref{tab:lffit}. The ionizing background produced
by AGNs has been compared with the values inferred
from the ionization status of the IGM, derived by different analysis of
the Lyman-$\alpha$ forest of high-z QSOs
\citep{fg08,calverley11,wyithe11,bb13,davies18,daloisio18}.

According to recent results by \citet{romano19}, the photo-ionization
rate produced by AGNs at $z\sim 4.6$ should be increased by a
factor of 1.7 with respect to the nominal value adopted in the
literature, due to a larger mean free path compared to
\citet{worseck14}. If the correction factor of 1.7
also yields at $z\sim 5.5$, then the photo-ionization rate in Table
\ref{tab:lffit} and Fig.\ref{fig:uvb}
turns out to be $\Gamma_{-12}=0.23^{+0.15}_{-0.20}$,
which is in agreement with the value of the UV background
measured at $z\sim 5.5$ by \citet{calverley11,davies18,daloisio18}.
Indeed, this value corresponds to 82\% of the UV background inferred by
\citet{davies18}, or alternatively to the 52\% of the one by \citet{daloisio18}.
A further correction of the $z\sim 6$ ionizing background of $+10\%$ could be
in place due to the impact of AGN jet lobes, according to \citet{torresalba20}.
These effects can have important consequences on the role of AGNs and
SFGs in the reionization of the Universe.

%-----------------------------------------------------------------

\section{Discussions}\label{sec:discussion}

The results of this paper rely on the AGN classification of two
objects, i.e. GDN3333 and GDS3073 in the CANDELS GOODS fields. The
former object is a secure AGN, due to the strong X-ray
emission by the 2Msec Chandra image
\citep{barger02,alexander03,giallongo19}. The nature of the latter
object, instead, has been questioned many times in the past
\citep{fontanot07,stark07,wiklind08,vanzella10,raiter10} with
different interpretations. We provide here clues in favor of the AGN
classification for GDS3073. We discuss in the following (Section
\ref{sub:agnvssfg}) the reliability of this classification. In
Section \ref{sub:reliable}, we discuss the implications of our
luminosity function determination.
In Section \ref{sub:whatnew} we discuss the new results we have achieved
from the analysis of the AGN luminosity function at $z>5$.

\subsection{Is GDS3073 an AGN or a SFG ?}\label{sub:agnvssfg}

The absence of any detectable X-ray emission \citep{giacconi02,giallongo19} or
radio production \citep{kellermann08} from GDS3073 down to relatively
low flux levels preliminarly suggests a SFG interpretation
for this object \citep{raiter10,vanzella10}. Composite galaxy spectra at $z\sim
3$ by \citet{shapley03} show high ionization emission lines (e.g. NV,
CIV) with P-Cygni profiles, which have been usually associated to
powerful stellar winds or shocks. This classification scheme has been also
adopted by \citet{heckman11} to interpret the spectra of local
galaxies with strong emission lines. On the basis of these
considerations, it was concluded that GDS3073 was a
(somehow peculiar) SFG. We discuss in the following
some evidences in favor of the AGN nature for this object.

GDS3073 was initially selected as an AGN candidate by
\citet{fontanot07} on the basis of the adopted optical color selection
and morphological criteria. The presence of an extremely strong
Lyman-$\alpha$ and a NIV] 1483,1486 inter-combination line in
emission, coupled with the apparent absence of NV 1240
emission line in the shallow FORS2 spectrum \citep{raiter10},
induced them to classify this object as an HII galaxy (similar to
the one described by Fosbury et al. 2003) at high-z.

The emission lines OVI 1032,1038, Lyman-$\beta$, Lyman-$\alpha$,
NV 1240, and NIV] 1483,1486 have been detected in the
X-Shooter, FORS2 or VIMOS spectra of GDS3073. The presence of these
high ionization lines in the UV rest-frame spectrum of GDS3073,
however, seems to indicate that this object is powered by an AGN
(see Fig. 5 by Le Fevre et al. 2019a). The powerful emission line
doublet NIV] 1483,1486, for example, has been usually
detected in high-redshift radio galaxies, as well as in rare optically
selected QSOs \citep{hainline11}. In addition, the NV
line at 1240 {\AA} rest-frame is an indicator of high ionization state
typical of narrow-line type II AGNs and low-luminosity QSOs
\citep{hainline11,laporte17}, and its presence indicates bona-fide AGN
radiation. The ionization potential of OVI is $>$100 eV, while
NV 1240 has a slightly lower threshold of $\sim$78 eV, much
more than expected from the typical emission of stellar populations
(see Fig. 1 of Feltre et al. 2016). The Lyman-$\beta$ 1026 {\AA} line
in emission is usually detected in QSOs/AGNs (see Fig. 14 by Le Fevre
et al. 2019a) but it is never detected in pure SFGs \citep{shapley03}.

According to \citet{chisholm19}, massive stars with age less than 3
Myr can produce non negligible NV 1240 emission, e.g. adopting the
BPASS stellar population synthesis model \citep{bpass}. In order to
have a prominent NV 1240 emission in the spectrum of a
SFG, the age of GDS3073 should be very low, below 5 Myr, since this
spectral feature drops very fast with the stellar population age and
it is observable only under rare circumstances. Given the red SED of
GDS3073, its high stellar mass and its present star formation rate, it
seems quite implausible that this object is a young starburst whose
light is dominated by very massive and extremely young stars. Indeed,
GDS3073 is strongly detected in the 5.8 and 8.0 $\mu$m IRAC bands,
where no nebular line or strong nebular continuum are expected at
$z=5.56$, suggesting a high galaxy stellar mass of $M_{star}\gtrsim
10^{10.7} M_\odot$. An AGN interpretation for GDS3073 is more natural,
and could explain in a simple way the presence of other strong
emission lines with high ionizing potential, such as NIV],
which is usually not observed in massive stars nor predicted by
stellar evolution models \citep{leitherer10,leitherer14}.
In addition, it is worth noting that \citet{glikman07} found a peculiar
faint QSO with NIV] 1483,1486 in emission and with P-Cygni profile for
NV 1240 emission line, similar to the GDS3073 spectral features,
and \citet{matsuoka19} reported NV 1240 P-Cygni emission for the stack of 18
narrow-line QSOs.
Thus, the P-Cygni profile is not a univocal signature of radiation by
massive stars, but it can be also associated to faint AGN activity.

The FWZI of the Lyman-$\alpha$ line for GDS3073 is of the order of
1050 km/s (Fig.\ref{fig:xshooter}). Such powerful outflows with
velocities exceeding 1000 km/s are beyond what stellar winds can
commonly produce. Strong outflows, detected in high ionization
absorption lines, are typically linked to the presence of hidden AGNs
\citep[e.g.,]{cimatti13,genzel14,karman14,brusa15,talia17,fs19,menci19}.
This is a further indication that the source
GDS3073 is being powered by an AGN.

The non detection of GDS3073 in X-ray is not
in contrast with the AGN classification for this source.
Steidel (2002) provided an example of a $z\sim 2$ AGN which was not
detected in the deep 1 Msec Chandra X-ray image of the GOODS-North
field. This object has not been detected in X-ray also in the deeper 2
Msec Chandra image, after adopting the X-ray forced photometry with
prior position from the HST detected sources, as described in
\citet{giallongo19}. It is thus not surprising that GDS3073 has not
been detected in the ultradeep 7 Msec Chandra X-ray image
\citep{giallongo19}. A similar case has been provided by \citet{mignoli19},
where they indicate that strong CIV emitters at
high-z are associated with bona-fide AGNs. In particular, the
CIV emitters with narrow lines have been classified as type-2
AGNs and they have only less than 50\% probability to be detected in
deep X-ray images. Recently, \citet{nardini19} report the discovery
of a significant sample (25\%) of X-ray under-luminous (by a factor of
3-10) QSOs among a well defined and homogeneous sample of 30 $z\sim 3$
QSOs that represent the most luminous, radio-quiet, non-broad
absorption line, and intrinsically blue quasar population. The
GDS3073 source in the CANDELS fields could be a similar case of X-ray
dim AGN at higher redshift and fainter luminosity.
It is worth noting at this aim that an X-ray obscuration is not
a clear indication of a negligible Lyman continuum
escape fraction, as discussed
in \citet{giallongo19}, since the absorption of UV and X-ray radiation are
originated by two distinct materials (hydrogen vs metals) surrounding the SMBH.

If we fit GDS3073 with stellar libraries such as \citet{bc03},
we obtain a star formation rate of 23
$M_\odot/yr$ and a stellar mass of $7\times 10^{10} M_\odot$
\citep{grazian15,santini15}. According to \citet{fs19}, there
is an elevated ($>30-50\%$) AGN fraction among compact
SFGs with $M_{star}>10^{10.7}M_\odot$ which are close to the galaxy
Main Sequence. Thus, it is likely that GDS3073 hosts an AGN at its
center, as shown in the previous sections through deep UV rest-frame
spectroscopy.

GDS3073 has been classified as an extended source in
\citet{vanzella10}, based on the comparison of its FWHM with observed
stellar profiles and on a morphological fitting using a
single Sersic profile as a template. We repeat a similar fit starting from a
deeper\footnote{Hubble Legacy Fields (HLF) Data Release 2.0 for the
GOODS-South region https://archive.stsci.edu/prepds/hlf/} HST F850LP
image \citep{hlf}, resampled at a pixel scale of 30 milliarcsec. We
use {\it galfit} version 3.0 \citep{galfit} with two components, i.e. a
point-like profile and a Sersic profile, leaving the centroid, the
magnitudes, and the morphological parameters of the Sersic profile
free to vary. In order to build the input PSF for {\it galfit}, we use
a sample of relatively bright stars within 2 arcminutes from GDS3073.
The resulting
best fit gives a magnitude for the point source of 24.75 in the F850LP
filter, while the extended source has a magnitude of 26.19 and a
Sersic index close to 1. The ratio between the two components is of the
order of 4, and it could resamble the ratio observed in local Seyfert
galaxies, where the host galaxy is clearly visible and is not outshined
by the active nucleus \citep[e.g.]{chen18}. If we convert the
apparent nuclear magnitude into absolute magnitude, we obtain
$M_{1450}=-21.87$, i.e. 0.23 magnitude fainter than the total absolute
magnitude inferred in Table \ref{tab:sample}. Adopting this
conservative value for our calculation of the AGN space density, the
$M_{1450}$ central bin moves by only 0.11 magnitudes, with negligible
effects on the parameterization of the luminosity function and the
derivation of the photo-ionization rate by AGNs.

We have provided here a number of observational evidences supporting
the presence of an AGN in GDS3073: the low-luminosity quasar/AGN
interpretation may explain the NIV] emission indicating the
presence of a hard radiation field, the broad Lyman-$\alpha$
component, the presence of other high ionization lines in emission
(OVI, Lyman-$\beta$, NV).
We can thus conclude that GDS3073 is a
robust AGN at $z\sim 5.5$.

\subsection{Is the AGN Luminosity Function at $z\sim 5.5$
reliable ?}\label{sub:reliable}

As already mentioned above, the estimate of the luminosity function in
Fig. \ref{fig:lfz5} is simply based on the ratio of the number of
spectroscopically confirmed AGNs at $5.0<z<6.1$ in the CANDELS fields
(GDN, GDS, and EGS) divided by the total cosmological volume of the
survey in the same redshift interval. No correction for incompleteness
have been applied to the observed space density.
It is worth noting here that an
extensive spectroscopic search for $z>5$ AGNs in the EGS field has not
been carried out yet, so in principle we could exclude the EGS field
from our calculations: if we follow this decision, the total survey
volume will be reduced by a factor of $\sim 1.6$, with a corresponding
enhancement of the space density by the same amount.

At high redshifts, a serious issue for the luminosity function
determination is the uncertainty due to the cosmic variance. We do
not take into account here the effect of cosmic variance on the error
bars of the luminosity function in Fig. \ref{fig:lfz5}. Since the
CANDELS fields adopted here are based on three extra-galactic areas in
different sky locations (GDN, GDS, and EGS), we expect that the
effect of cosmic variance is mitigated.

\subsection{What can we learn from the AGN Luminosity Function
at $z\sim 5.5$ ?}\label{sub:whatnew}

The AGN nature of GDN3333 and GDS3073 is of particular importance, since strong
claims of the lack of numerous high-z AGNs at the faint side of the UV
luminosity function \citep[e.g.]{matsuoka18,mcgreer18} are based on the lack
of objects showing broad emission lines. These lines are indeed
characteristics in bright QSOs typically detected at bright absolute
magnitudes of $M_{1450}\sim -28$. The narrow Lyman-$\alpha$ lines of
GDN3333 and GDS3073 indicate instead that the search for faint high-z AGNs must
be carried out in a different way, for example through deep
spectroscopy from space, e.g. with JWST.
Ultra-deep spectroscopy from the ground could be a viable alternative
before the advent of JWST, provided that the exposure time will be
long enough, as done for GDS3073, or the observations will be carried out with
powerful future generation telescopes (e.g. GSMT, EELT, TMT).
Interestingly, recent results by \citet{matsuoka19} indicate that
$\sim 20\%$ of QSOs at $z>5.7$ are showing strong and narrow
Lyman-$\alpha$ in emission. Their spectra are clearly different from
that of the other QSOs, and show also NV 1240 in emission.
These narrow line QSOs could be even more common at faint
magnitudes, in the luminosity regime covered by the GDN3333 and
GDS3073 sources.

The luminosity functions of \citet{matsuoka18,mcgreer18}
are significantly underestimated w.r.t. our
determination in CANDELS, which, it is worth stressing, is not subject
to any bias or systematics. As discussed in \citet{boutsia18} and
\citet{giallongo19}, optical surveys based on optical color selection can be
subject to strong incompleteness. This is especially true if their
efficiencies are relatively high at $z>5$, as for the case of
e.g. \citet{matsuoka18}. This could indicate that their selection
criteria are rather stringent, in order to avoid the many
contaminants, but their completeness turns out to be very low. Indeed, the
color selection adopted by \citet{matsuoka18} is rather stringent, of
$i-z>2.0$. If we check the optical photometry of GDN3333 and GDS3073
from CANDELS, we obtain much bluer $i-z$ colors, of 0.12 for GDN3333
($I_{775}=24.03$ and $Z_{850}=23.91$) and 0.69 for GDS3073 ($I_{775}=25.22$ and
$Z_{850}=24.52$). Adopting the \citet{matsuoka18} color criterion, these
two sources would not have been selected as high-z AGNs. It is thus
not surprising at all the low level of the AGN space densities observed by
\citet{onoue17,matsuoka18,mcgreer18} and \citet{shin20} with respect to
our determination.

We note that our criterion does not rely on any limitation in terms of
line width, consequently it is not biased with respect to the standard
AGN spectral classification. Indeed, both type-1 and type-2 AGNs can
be selected by our criterion. GDS3073 has a Lyman-$\alpha$ line
width of 600 km/s, or somewhat larger due to the strong IGM absorption
along the blue wing \citep{vanzella10}. This value is typical of
type-2 narrow-lined AGNs whose line widths are in the range $500 -
1000$ km/s. Typical faint narrow-lined AGNs similar to our source are
also present in the large LBG sample by \citet{steidel02}.
The presence of OVI emission along the line of sight shows the
presence of highly ionized gas by hard UV photons. It is to note in
this respect that recent results on type-2 AGNs are suggesting that
geometrical/obscuration effects are not the main mechanisms producing
the spectral differences between type-2 and type-1 AGNs. Indeed,
type-2 AGNs often show faint broad components associated with average
smaller black hole masses for a given X-ray luminosity
\citep{onori17}. These considerations support the hypothesis of a
significant contribution by this sub-class of AGNs whose abundance in
the early phases of galaxy formation at $z>5$ is just emerging from
our spectra and similar observations in other surveys.

Another consideration is that the incompleteness corrections applied
by \citet{giallongo19} relied on the {\it observed} distribution of
AGN candidates in the UV-rest/X-ray plane since their intrinsic
distribution is not known a priori, especially at $z>5$. For this
reason, the UV/X-ray flux ratio could be biased towards brighter X-ray
emitters, and thus the completeness correction in the volume densities
of X-ray selected sources should be considered as a conservative lower
limit, as discussed in \citet{giallongo19}. The fact that GDS3073 is a
confirmed AGN without X-ray emission may indicate that the
incompleteness in \citet{giallongo19} is of the order of 50\% at
$M_{1450}=-22$.

These limitations can in general be alleviated at shallower UV-rest
and X-ray flux limits over larger sky areas. The progressive reduction
in the last years of the Chandra sensitivity in the soft energy
bands, however, prevents the planning of future medium-deep surveys
in such larger areas.
A multiwavelength effort is required to increase our knowledge on the
space density of AGNs at $z>5$ and intermediate UV luminosities.
Our paper shows that detailed spectral analysis
focused on the detection of the NV and OVI high ionization emission
lines in very deep spectra (with JWST and ELT in the future) of the UV
brightest, Lyman-$\alpha$ emitters can be an effective method to
reveal signs of AGN activity in bright $z>5$ star forming galaxies.
A further complementary approach comes from variability analysis in
deep multicolor fields. In a recent study by \citet{pouliasis19},
source variability has been investigated in the GOODS-South region
providing the selection of faint AGNs up to $z\sim 5$. Noticeably,
only 26 out of these 113 AGNs (23\%) have been detected in X-ray by
the ultra-deep 7 Msec Chandra image, indicating that there is a
possible incompleteness of X-ray selected AGNs at high-z and faint
luminosities. In \citet{giallongo19} we have found similar correction
factors of $\sim$2-3 at $M_{1450}\sim -20$, confirming these results.

An unbiased determination of the $z>5$ AGN luminosity function is
useful in order to interpret correctly the number density of
SFGs at high-z. The AGNs studied in this paper have
UV luminosities which are comparable or fainter than the brightest
SFGs observed at these redshifts \citep{finkelstein19}.
It is thus possible that some AGNs brighter than the
characteristic magnitude ($M_{1450}\sim$-20) of the LBG luminosity
function at $z\sim 5-6$ can contribute to enhance its normalization
artificially. The case of object GDS3073 is anecdotal. In previous
works it has been classified as a SFG, and it thus
contributed to the bright side of the LBG luminosity function at
$z>5$. As also proposed by \citet{kulkarni19}, it is possible
that an enhanced incidence of faint AGNs at high-z is consistent with
the flat bright-end slopes detected by
\citet{bowler12,bowler14,bradley14,bowler15}
for the $z\sim 7$ UV
luminosity function of galaxies relative to a Schechter function. At
lower redshifts, i.e. $z\sim 4$, the issue of the transition region
between bright galaxies and faint AGNs has been studied by
\citet{stevens18,adams19}. Based on the recent results of
\citet{boutsia18}, \citet{adams19} concluded that there is a
significant contribution of AGNs in the bright side of the luminosity
function of galaxies, at $M_{UV}\sim -23$, confirming our hypothesis.

The AGN nature of GDN3333 and GDS3073 is also important for the
interpretation of the ratio of Lyman-$\alpha$ emitters among LBGs at
$z>6$ and its implications for reionization \citep{hoag19}. Both these
AGNs, and in particular GDS3073, could have been misidentified as
bright LAEs at high redshift, systematically biasing the observed
LAE/LBG ratio towards higher values. As a notable example at higher
redshifts, the peculiar LAEs spectroscopically confirmed, e.g., by
\citet{stark15,oesch15,borsani16,schmidt16,laporte17,mainali18,tilvi20} at
$z>7$ are not expected due to an almost neutral IGM
\citep{keating19}. The Lyman-$\alpha$ detections
could instead indicate that these objects are
powered by highly ionizing radiation not produced by
star-formation, and they are able to carve large ionized bubbles.
Indeed, this hypothesis is corroborated by the
detection of high-ionization lines, e.g. NV, CIV,
HeII, OIII, CIII, in their UV rest-frame spectra.

\subsection{The IGM temperature and the HeII reionization in an
AGN-driven reionization scenario}\label{sub:tigm}

Two issues related to an AGN-driven reionization scenario are a too
high IGM temperature predicted at $z>2$ compared to the Lyman forest
estimates \citep{bb13} and a too early HeII reionization expected
at $z>4$, which
according to observations should be almost completed at $z\lesssim 3$
\citep{garaldi19}.
Recent measurements have revised upward the IGM temperature measured
in the Lyman forest at $z\sim 2-4$ \citep{hiss18,walther18,hiss19,telikova19}
and at $z=5.4-5.8$ \citep{gaikwad20}. In particular the latter work
confirms the late and inhomogeneous scenario, where islands
of neutral material are still persistent till $z\sim 5.2-5.3$, with
strong spatial fluctuations of the IGM temperature.

An early population of faint AGNs can sustain extended HeII
reionization at
$z>4$, but predicts a too low optical depth in HeII at $2.7<z<3.0$
\citep{puchwein18}. Using high S/N ratio QSO spectra by COS-HST,
\citet{worseck18} observed $\tau_{HeII}\gtrsim 4$ at $z\sim 3.8$ and
narrow transmission spikes in the HeII forest at $z>3.5$, the signposts
of isolated patches of fully reionized helium. These measurements are
in agreement with the predictions based on the stochasticity of the
space density of ionizing sources such as high-z AGNs and the partial
reionization of optically thick absorbers near the completion of the
reionization process \citep{compostella14,chardin15,chardin17,madau17}.
In practice, strong fluctuations in HeII
absorption could be due to spatial
variation of the hard UV background in an already ionized medium at $z\sim 3$
\citep{morrison19}.
A soft UV background produced by a homogeneous distribution of ubiquitous
ultra-faint SFGs at $z>10$ is in tension with such
measurements. Moreover, if the hydrogen reionization is driven by
SFGs with hard ionizing radiation, that show strong
HeII, NV or CIV emission lines \citep{schaerer19,chisholm19}, they should
emit copious
amount of high-energy photons with $E>54.4$ eV, with the drawback,
highlighted above, of a too early HeII reionization at $z>6$.

%-----------------------------------------------------------------

\section{Summary and Conclusions}\label{sec:conclusion}

We have used deep HST and ground-based images from the CANDELS
GOODS-North, GOODS-South, and EGS fields in order to derive the space
density of faint ($L\sim 0.1L^*$, M$_{1450}\sim -22.5$) AGNs at $z>5$.
Thanks to the deep VLT spectroscopy with FORS2, VIMOS, and X-Shooter,
we confirm the AGN nature for the source GDS3073 in the GOODS-South
field, at $z_{spec}=5.563$ and M$_{1450}=-22.1$. The
presence of high-ionization
emission lines (OVI, Lyman-$\beta$, NV, NIV])
detected in the UV rest-frame spectra of GDS3073 indeed corroborates
its AGN activity, even if this source is not
detected in the ultradeep 7 Msec X-ray image by Chandra.
The deep 2 Msec Chandra X-ray images of the GOODS-North field allow us
to select another AGN, GDN3333, at $z_{spec}=5.186$ and
M$_{1450}=-22.6$. Interestingly, without the availability of deep
Chandra X-ray imaging, this object would be classified as a normal
SFG, due to the limited spectroscopic information
available in the UV rest-frame wavelengths.

The AGN space density at M$_{1450}\sim -22.5$ has been computed by
dividing the number of observed AGNs by the total cosmological volume
within $5.0<z<6.1$ in the CANDELS footprints (area=551.5 sq. arcmin.),
without any correction for incompleteness. The derived space density
of faint AGNs in the redshift interval $5.0<z<6.1$ is a lower
limit for two main reasons: 1-we do not correct our data for
incompleteness, which could be present at such faint observed
magnitudes. 2-The adopted CANDELS area comprises also the EGS field,
but the spectroscopic search for high-z objects in EGS has not been
carried out as extensively as in the GOODS fields, thus a reduction of
the effective survey area (of a factor 1.6) could be applied, with a
corresponding enhancement of the luminosity function by the same amount.

We obtain a space density of $\Phi=1.29\times 10^{-6} cMpc^{-3}$ at
$z\sim 5.5$ and M$_{1450}\sim -22.3$ ($L\sim 0.1L^*$). This value is
much higher than recent determinations in the literature
\citep{mcgreer18,matsuoka18,cowie20}, which could
be affected by severe incompleteness, and it is marginally
consistent with
\citet{marchesi16,ricci17,onoue17,parsa18,kulkarni19,shen20,shin20}.
Our new value is consistent or even
higher than the one derived by \citet{giallongo19}, although statistics
in our measurements
are still poor and could be affected by cosmic variance.

Connecting our new point at M$_{1450}\sim -22.3$ with other AGN volume
densities both at brighter \citep{jiang16,yang16,chehade18} and
fainter magnitudes as described in \citet{giallongo19}, we have
derived a faint-end slope of $\beta\sim 1.5$. This slope is similar to
the one derived by \citet{giallongo19} at $z\sim 4.5$, although
slightly flatter. The new determination of the AGN luminosity function
in this paper, together with the results derived by
\citet{giallongo19}, can provide new predictions on the redshift
evolution of the global AGN number density at $z>6$, revising previous
expectations \citep[e.g.]{kulkarni19,shen20}.

A Photo-ionization rate $\Gamma_{-12}=0.13$ has been derived by
adopting our new best-fit of the AGN luminosity function and standard
redshift evolution of the mean free path of ionizing photons into the
IGM \citep{worseck14}. We note in this context that adopting the new
evolutionary scenario for the mean free path derived by
\citet{romano19}, would result in a photo ionization rate $\sim 1.7$
times higher, yielding $\Gamma_{-12}=0.23$. This value for the
HI ionizing background is in agreement ($\sim 52-82\%$)
to the recent estimates at
$z\sim 5.5$ derived from the analysis of the ionization level of the
IGM \citep{calverley11,davies18,daloisio18} and suggests an important
or even dominant contribution of the global AGN population to the
ionizing UV background into the reionization epoch at $z=6-7$, if such trend is
confirmed at higher redshifts.

An important role of AGNs and QSOs in the cosmological reionization
process may also provide a natural explanation for the large
line-of-sight scatter on scales of 50 cMpc in the Lyman-$\alpha$
opacity at $z=6$ \citep{becker15,chardin15,bosman18,eilers18,keating19,
bosman20,meiksin20} and for the quick and late drop of the
neutral fraction of the IGM from $z=7.5$ to $z\lesssim 6$
\citet{hoag19,keating19,yung20}.
Such features are not expected in galaxy-dominated reionization
models \citep{naidu19}.

As discussed in Section \ref{sub:vimos}, deep spectroscopic follow-up
is required in order to determine the nature (SFG vs AGN) of high
redshift sources selected by means of multi-wavelength criteria. In
particular, deep spectroscopy is fundamental to detect the faint
NV 1240 or CIV 1549 emission lines, which are clear
signatures of the AGN activity even in narrow emission line sources
\citep{boutsia18,matsuoka19}. Confirmation of $z>5$ AGNs requires ultra-deep
spectroscopy extended in the NIR regime, on ground based 8-10m class
telescopes. Indeed, the failure in the detection of NV in
emission in some of the brightest $z>6$ galaxies could be simply due
to the currently adopted shallow UV rest-frame spectra \citep[e.g.]{capak11}.

An updated determination of the AGN luminosity function is particularly
helpful for a correct evaluation of the SFG space
density around $M_{UV}\sim -22$. If the galaxy sample is contaminated by faint
AGNs, which are mimicking the presence of bright and rare SFGs, it is
possible to overestimate the galaxy luminosity function at the bright
end. It is thus possible that the results by
\citet{bowler12,bowler14,bradley14,bowler15}
at $z\sim 7$ or by \citet{bridge19} at $z\sim 8$ are
affected by the presence of hidden faint AGNs. Moreover, cleaning the
SFG samples from sources powered by AGNs is
fundamental in order to measure with high accuracy the Lyman continuum
photon production efficiency. Present renditions of $\xi_{ion}$ at
lower redshifts \citep{nanayakkara20}, indeed, could be biased high
due to the presence of unrecognized AGNs, especially the sources
dominated by high ionization lines in emission, e.g. those
characterized by high OIII/OII line ratio or large
EWs in HeII or H$\alpha$. A clear separation between
SFGs and AGNs is also important in order to derive an
unbiased estimate of the fraction of LAEs among LBGs, which have deep
implications on the derivation of the neutral hydrogen fraction at
high-z.

Deep JWST spectroscopy with NIRSpec of all the known Lyman-$\alpha$
emitters and LBGs at $z>6$ down to luminosities of $\sim 0.1L^*$ will
be of fundamental importance to reveal the abundance of faint AGNs at the
EoR. In the future GMT, EELT, and TMT telescopes
will shed light on the early accretion history in the Universe. In
particular, the AO-assisted MICADO instrument \citep{micado} at the
ESO ELT will provide the most detailed morphological information of
the continuum and nebular emission from sources hosting young
accreting SMBHs.

%-----------------------------------------------------------------

\acknowledgments
We thank the referee for the constructing report, which
has been very useful to improve the overall quality of the present paper.
This work is based on data products from observations made with ESO
Telescopes at La Silla Paranal Observatory under ESO programmes ID
384.A-0886(A), 089.A-0679(A), 170.A-0788, 194.A-2003(E-Q). Based on
observations made at the European Southern Observatory, Paranal, Chile
(ESO programme 170.A-0788) The Great Observatories Origins Deep
Survey: ESO Public Observations of the SIRTF Legacy/HST
Treasury/Chandra Deep Field South); on observations obtained with the
NASA/ESA Hubble Space Telescope obtained at the Space Telescope
Science Institute, which is operated by the Association of
Universities for Research in Astronomy (AURA), Inc.; and on
observations made with the Spitzer Space Telescope, which is operated
by the Jet Propulsion Laboratory, California Institute of Technology
under a contract with NASA. AG and FF acknowledge support from PRIN
MIUR project 2017-PH3WAT ``Black Hole winds and the Baryon Life Cycle
of Galaxies: the stone-guest at the galaxy evolution supper''.

%% To help institutions obtain information on the effectiveness of their 
%% telescopes the AAS Journals has created a group of keywords for telescope 
%% facilities.
%
%% Following the acknowledgments section, use the following syntax and the
%% \facility{} or \facilities{} macros to list the keywords of facilities used 
%% in the research for the paper.  Each keyword is check against the master 
%% list during copy editing.  Individual instruments can be provided in 
%% parentheses, after the keyword, but they are not verified.

%\vspace{5mm}
%\facilities{HST(WFC3), CXO, VLT}

%% Similar to \facility{}, there is the optional \software command to allow 
%% authors a place to specify which programs were used during the creation of 
%% the manuscript. Authors should list each code and include either a
%% citation or url to the code inside ()s when available.

%\software{}

%% For this sample we use BibTeX plus aasjournals.bst to generate the
%% the bibliography. The sample63.bib file was populated from ADS. To
%% get the citations to show in the compiled file do the following:
%%
%% pdflatex sample63.tex
%% bibtext sample63
%% pdflatex sample63.tex
%% pdflatex sample63.tex

\bibliography{sample63}{}
\bibliographystyle{aasjournal}

%% This command is needed to show the entire author+affiliation list when
%% the collaboration and author truncation commands are used.  It has to
%% go at the end of the manuscript.
%\allauthors

%% Include this line if you are using the \added, \replaced, \deleted
%% commands to see a summary list of all changes at the end of the article.
%\listofchanges

\end{document}